\documentclass[preprint,showpacs,preprintnumbers]{revtex4}

\usepackage{graphicx}
\usepackage{amsmath}

\begin{document}

\title{Artificial molecular quantum rings: Spin density functional theory calculations}
\author{L. K. Castelano$^a$}
\author{G.-Q. Hai$^a$}
\email{hai@ifsc.usp.br}
\author{ B. Partoens$^b$}
\author{F. M. Peeters$^b$}
\email{francois.peeters@ua.ac.be}
\affiliation{$^a$Instituto de F\'isica de S\~ao Carlos, Universidade de S\~ao Paulo,
13560-970, S\~ao Carlos, SP, Brazil \\
$^b$Department of Physics, University of Antwerp, Groenenborgerlaan 171,
B-2020 Antwerp, Belgium}

\begin{abstract}
The ground states of artificial molecules made of two vertically coupled
quantum rings are studied within the spin density functional theory for
systems containing up to 13 electrons. Quantum tunneling effects on the
electronic structure of the coupled rings are analyzed. For small ring
radius, our results recover those of coupled quantum dots. For intermediate
and large ring radius, new phases are found showing the formation of new
diatomic artificial ring molecules. Our results also show that the tunneling
induced phase transitions in the coupled rings occur at much smaller
tunneling energy as compared to those for coupled quantum dot systems.
\end{abstract}

\maketitle

\DeclareGraphicsExtensions{.jpg, .pdf, .mps, .png, .tiff}

\section{Introduction}
The realization of electronic structures by design is one of the ultimate
goals of nano-science. By varying the size and geometry of
nanocrystallites,\ it is possible to tune and control the confined quantum
states and thus the electronic, magnetic and optical properties. A very
successful example of such systems are semiconductor quantum dots and
molecules. These artificial atoms and molecules\cite{reimann} exhibit
shell-filling and new molecular many-body states which have been confirmed
experimentally. These structures can be used as more efficient lasers and as
hosts for storing quantum information, \textit{i.e}. qubit, because of their
atomic-like properties.

Quantum rings (QRs) are known for the Aharonov-Bohm effect and its
persistent current\cite{AB} where the phase rigidity of the electron wave
leads to quantum\ oscillations in the current. Such nanometer-sized rings
are artificial benzene analogues. Experimentally there have been several
approaches to fabricate ring structures ranging from top-down approaches,
\textit{e.g}.\ nano-lithography (\textit{e.g}. atomic force microscope
patterning)\cite{held}, to bottom-up approaches as strain induced
self-organization\cite{garcia} and droplet MBE epitaxy\cite{gong}. The
bottom-up approach has been used recently for nano-structured assembly of
quantum ring complexes as double rings\cite{mano} and vertically coupled
rings\cite{granados}. Due to the ring geometry, these ring complexes open a
new route for measurement of quantum interference effects and for novel
many-body states.

In the present work we study two vertically coupled rings which
exhibit new molecular many-body states, hitherto not found in other
systems. The tunability of the ring radius and the inter-ring
distance leads to a very rich variety of many-electron ground
states. In previous theoretical studies for single rings, Simonin
\textit{et al.}\cite{simo} showed that a displaced parabolic model
is suitable for describing the confinement potential for a
self-assembled semiconductor QR. Many-body spectra of single QRs
have been obtained using exact diagonalization techniques\cite{kos}
and spin-density-functional theory\cite{lin}. A study of the
persistent current in QRs for up to 6 electrons showed that the
current-spin-density-functional theory provides a suitable tool for
describing the physics related to the Aharonov-Bohm
effect\cite{vief}. Very recently, theoretical studies were
reported on the single-electron spectrum in two vertically coupled QRs\cite%
{li,cli} and of few electron eigenstates of concentric double QRs\cite{sza}.

Two vertically coupled quantum rings (CQRs) form a new type of artificial
molecule (AM) where the ring radius together with the inter-ring distance
are new tunable parameters providing new degrees of freedom to modulate and
control the electronic structure of the artificial ring shaped \
\textquotedblleft molecules\textquotedblright . In the present work, we
apply the well established spin-density functional theory to study the
ground state configurations or phases of few-electron CQRs. Full phase
diagrams of these new quantum ring AMs are obtained up to 13 electrons.

\section{Theoretical model}
The electrons are confined in a plane where the confinement
potential is modeled by a displaced parabolic function
$V(r)=\frac{1}{2}m^{\ast }\omega _{0}^{2}(r-r_{0})^{2}$, where
$\mathbf{r}=(x,y)=(r,\theta )$, $\omega _{0}$ is the confinement
frequency and $r_{0}$ is the radius of the ring. The tunneling
between the two stacked identical rings in the $z$ direction occurs
through a potential $V(z)$ modeled by two coupled symmetric GaAs
quantum wells with a finite barrier height $V_{0}=250$ meV and a
well width of $W=120$ \AA . For these parameters we obtain for the
energy splitting $\Delta =22.86$ $\exp [-d($\AA\ $)/13.455]$ meV
between the bonding and anti-bonding states when the two wells are
separated by a distance $d$.

We use the Kohn-Sham orbitals $\psi _{nm\sigma }(\mathbf{r})=\exp (-im\theta
)\phi _{nm\sigma }(r)Z(z)$ to express the density and ground state energy.
The eigenfunctions $\phi _{nm\sigma }(r)$ are expanded in the Fock-Darwin
basis to solve the Kohn-Sham equation. The contribution from higher excited
states is neglected because the confinement in the $z$ direction is much
stronger than that in the plane. Therefore the motion in the $z$ direction
may be assumed to be decoupled from the in-plane motion. The total density
in the rings is $\rho (r)=\sum_{\sigma }\sum_{n,m}|\phi _{nm\sigma }(r)|^{2}$
where $\sigma =(\uparrow ,\downarrow )$ being the $z$ component of the
electron spin. Following Ref.\cite{bart}, we approximate the density in the $%
z$ direction by $\delta $ functions at the center of the quantum wells. The
exchange-correlation energy is treated within the local density
approximation using the Tanatar-Ceperley \cite{tan} functional. The ground
state energy and the corresponding electronic configuration of the system
are determined by comparing the total energies of all possible
configurations. The ground state phases are labeled by three quantum numbers
$(S_{z},M_{z},I_{z})$: total spin $S_{z}$, total angular momentum $M_{z}$
and the isospin quantum number $I_{z}$, which is the difference between the
number of electrons in the bonding state and in the antibonding state
divided by 2. For the CQRs, the ground state phases are found for different
inter-ring distances (tunneling energy $\Delta $) and ring radii. In the
limits of small and large inter-ring distance $d$, the results for single QRs%
\cite{lin} are recovered and in the limit of small ring radius ($%
r_{0}\rightarrow 0$), the previous results for two coupled quantum
dots (CQDs)\cite{bart,pi} are obtained. The energy levels $\epsilon
_{n,m}$ of a single quantum ring are shown in Fig. 1 as a function
of the ring radius (in unit a$_{0}=\sqrt{\hbar /m^{\ast }\omega
_{0}}$). The energy levels are labeled by the radial quantum number
$n=0,1,2,...,$ and the angular quantum number $m=0,\pm 1,\pm 2,...$.
As compared to the quantum dot case, in a
ring the degeneracy of, \textit{e.g.} the states $\epsilon _{0,\pm 2}$ and $%
\epsilon _{1,0}$ ($\epsilon _{0,\pm 3}$ and $\epsilon _{1,1}$, \textit{etc}%
.) is lifted. With increasing ring radius, the energy difference $\epsilon
_{1,0}-\epsilon _{0,\pm 2}$ increases and tends to a constant $\hbar \omega
_{0}$ at large $r_{0}$. The effect of lifting the degeneracy becomes
significant for systems with 9 electrons or more when the energy levels $%
\epsilon _{1,0}$ and $\epsilon _{0,\pm 2}$ become occupied. Notice
that the low lying energy levels of the $m=0$ orbitals exhibit a
minimum as a function of the ring radius. This effect can be
understood qualitatively by looking at the potential height $m^{\ast
}\omega _{0}^{2}r_{0}^{2}/2$ in the center of the ring that is
indicated by the thick dot-dashed curve in Fig. 1. On the left side
of this curve, the system behaves similar to a quantum dot and to
the right of it, a quantum ring starts to form. The local minimum in
the $m=0$ curves occurs at this crossover regime and will affect the
ground state phases for CQRs with 12 and 13 electrons with
intermediate ring radius.

We want to stress that the above theory is based on the envelope
function approach and the local-spin density approximation (LSDA).
These approximations are valid in the present study for the coupled
quantum rings. For a confinement strength of $\hbar\omega_0$ = 5
meV,  the corresponding typical length scale is a$_0 \simeq $ 15 nm
in the case of GaAs\cite{held}. We show results for a diameter up to
2$\times$ 5$\times$a$_0$ = 150 nm. These scales are still much
larger than the interatomic distances, therefore the effective mass
approximation is valid. Moreover, we have chosen for a LSDA-DFT
approach to study the many-electron states of the CQRs. This
approach was used before successfully to describe the energy levels
in a single quantum dot of comparable size\cite{LDA} and worked
surprisingly well even for systems of 2 and 3 electrons. Ferconi and
Vignale\cite{LDA} compared the exact result for the ground state
energy of a two electron quantum dot with a DFT calculation. The
accuracy is found to be better than $3\%$. Therefore we believe that
claims we make that are based on the LSDA-DFT approach are
justified. We did not implement a higher functional, because it
would not reveal new physics. Our confinement potential is anyhow a
model system for two coupled rings.

\section{Numerical results and discussions}
The stability region of the different ground states are summarized
in phase diagrams in the $\Delta $ versus $r_{0}$ plane for systems
with fixed electron number $N$ (see Figs. 2(a-d)). The insets in the
figures give the single-particle picture representation for each
phase. In the calculations, we took the confinement energy $\hbar
\omega _{0}=5$ meV. Several of the transitions between the phases
can be understood from a single-particle (SP) picture when the
exchange-correlation effects are added. The confinement in the $z$
direction splits the single QR levels (Fig. 1) into a set of bonding
and antibonding levels by $\Delta $.

>From such a SP picture, one does not expect configurations different
from the CQDs as long as the level $\epsilon _{1,0}$ is not
occupied. Indeed, for CQRs with $N\leq 8$ we found qualitatively
similar ground states as for CQDs. For $N=8$, four different phases
are found as shown in Fig. 2(a). For
fixed $r_{0}$, there is a single ring atomic type phase $(2,0,0)$ at small $%
\Delta $ where the inter-ring distance is large and the rings are decoupled.
For large $\Delta $ the two rings are strongly coupled and act as a single
one resulting in another atomic type phase $(1,0,4)$. There are two
molecular type phases: $(0,0,2)$ and $(1,2,3)$ for intermediate inter-ring
distance similar to those found earlier for CQDs. It is clear that, with
decreasing $\Delta $, more electrons from the antibonding state transfer to
the bonding state leading to phase transitions. Because the SP levels $%
\epsilon _{0,m}$ (for $m=0$, $\pm $1, and $\pm $2) decrease with increasing
ring radius $r_{0}$ and converge at large ring radius, the atomic type phase
$(1,0,4)$ of strongly coupled QRs is enhanced at large $r_{0}$. To this end,
we can also understand that the tunneling effect is enhanced at large $r_{0}$%
. This becomes obvious when we investigate the phase transitions with
increasing $r_{0}$ for a fixed $\Delta $. At $\Delta =2$ meV ($d=$ 32.8\AA ),%
\textit{\ e.g}., the transitions $(0,0,2)\rightarrow (1,2,3)\rightarrow
(1,0,4)$ corresponds to the isospin transitions $I_{z}=2\rightarrow
3\rightarrow 4$ where the maximum value is reached for large $r_{0}$. Notice
that the magnetic state, \textit{i.e}. the $M_{z}$ value, can be manipulated
by varying $\Delta $ and/or $r_{0}$.

Novel phases appear when $N\geq 9$. Let's first study the case of\ 9
electrons (see Fig. 2(b). The phase $(3/2,0,9/2)$, for $r_{0}<$ 0.89 a$_{0}$%
, is consistent with the result obtained for CQDs\cite{bart}. A ring atomic
type phase $(1/2,2,9/2)$ appears for $r_{0}>$ 0.89 a$_{0}$ in the strong
tunneling regime. The emergence of the new phase $(1/2,2,9/2)$ originates
from the lifting of the degeneracy of the SP levels $\epsilon _{1,0}$ and $%
\epsilon _{0,\pm 2}$ in the bonding states. When the ring radius increases, $%
\epsilon _{1,0}-\epsilon _{0,\pm 2}$ increases and when this energy
difference overcomes the exchange energy due to Hund's rule, the state $%
\epsilon _{0,\pm 2}$ becomes energetically more favorable than $\epsilon
_{1,0}$. Consequently, a ground state transition undergoes from $(3/2,0,9/2)$
to $(1/2,2,9/2)$.

Another new result for the CQRs of 9 electrons is the disappearance of the
ground state $(1/2,2,5/2)$ for $r_{0}>$ 3.04 a$_{0}$ which can be understood
as follows: (i) the energy difference $\epsilon _{0,\pm 1}-\epsilon _{0,0}$
in the antibonding states is small at large $r_{0}$ (it is much smaller than
$\epsilon _{0,\pm 2}-\epsilon _{0,\pm 1}$); and (ii) the existence of the $%
(3/2,0,7/2)$ phase is largely due to the exchange gain of the two aligned
spin states in $\epsilon _{0,-2}$ and $\epsilon _{0,+2}$ in the bonding
states. For $r_{0}>$ 3.04 a$_{0}$, this exchange gain becomes larger than $%
\epsilon _{0,\pm 1}-\epsilon _{0,0}$. With decreasing $\Delta $, as soon as
one of the electrons in the $\epsilon _{0,\pm 2}$ levels in the bonding
states jumps to the antibonding state, the other electron follows leading to
the transition from $(3/2,0,7/2)$ directly to the $(1/2,1,3/2)$ phase. This
is a clear manifestation of the more pronounced effect of the
electron-electron interaction in CQRs.

For 11 (and 12) electrons the number of possible ground states jumps to 8
(see Fig. 2(c)). Among them, the ring atomic type phase $(1/2,3,11/2)$ and
the ring molecular type phase $(1/2,0,9/2)$ appear only at finite ring
radius. The transition from $(1/2,0,11/2)$ to $(1/2,3,11/2)$ at $r_{0}=$
2.05 a$_{0}$ is associated with the transfer of a bonding state electron
from $\epsilon _{1,0}$ to $\epsilon _{0,\pm 3}$. This is almost completely
determined by the SP energy crossover of the two levels at $r_{0}=$ 2.07 a$%
_{0}$. We also notice that the molecular type phase $(1/2,2,7/2)$ is
completely suppressed\ at large ring radius ( $r_{0}>$ 3.17 $a_{0})$. A
similar phenomenon was already found for the $N=9$ case and is a consequence
of the fact that the energy difference $\epsilon _{0,\pm 1}-\epsilon _{0,0}$
in the antibonding state at large $r_{0}$ becomes smaller than the exchange
gain in the $(3/2,1,5/2)$ phase. We also observe a stabilization of the ring
molecular type phase $(1/2,0,9/2)$ at intermediate ring radius where the
inter-ring tunneling energy $\Delta $ leads to a crossing of the lowest
level $\epsilon _{0,0}$ in the antibonding state with the level $\epsilon
_{0,\pm 3}$ in the bonding state. The lowest antibonding level $\epsilon
_{0,0}$ is almost flat with increasing $r_{0}$ while the level $\epsilon
_{0,\pm 3}$ decreases rapidly. As a consequence, the phase $(1/2,0,9/2)$
with one electron in the antibonding level $\epsilon _{0,0}$ is enhanced.

The number of possible phases increases to 13 for a CQR with 13 electrons
(see Fig. 2(d)). From left to right on the top of the figure, we see 3
phases: $(1/2,3,13/2)$, $(3/2,0,13/2)$, and $(1/2,3,13/2)^{\ast }$ where the
star denotes that, in this new ring type phase, the levels $\epsilon _{0,\pm
3}$ of higher angular momentum are occupied instead of the level $\epsilon
_{1,0}$ in the quantum dot type phase $(1/2,3,13/2)$. The phase $%
(1/2,3,13/2)^{\ast }$ is due to the SP level $\epsilon _{0,\pm 3}$ which
becomes lower in energy than the level $\epsilon _{1,0}$ in a quantum ring.
The intermediate phase $(3/2,0,13/2)$ is due to many-body
exchange-correlation effects. With decreasing $\Delta $, one of the
electrons in the bonding state in the above configurations jumps to the
lowest antibonding level leading to three molecular type phases: $%
(1/2,0,11/2)$, $(3/2,3,11/2)$, and $(3/2,0,11/2)$.

With further reducing the inter-ring tunneling energy, transitions occur
from $(5/2,0,5/2)$ to $(3/2,2,5/2)$ at $r_{0}=$ 0.99 a$_{0}$\ and to $%
(1/2,1,7/2)$ with increasing ring radius. These ring molecular type phases
are a consequence of the lifting of the degeneracy of the levels $\epsilon
_{0,\pm 2}$ and $\epsilon _{1,0}$ at finite ring radius and results also in
a more stable $(3/2,1,3/2)$ phase. The molecular type phase $(1/2,3,9/2)$
turns to be unstable for $r_{0}>3.89$ a$_{0}$ due to a strong
exchange-correlation gain in its neighbor phase $(3/2,0,11/2)$. For CQRs of
13 electrons, we found that the following transitions are invoked by state
changes of two electrons: $(1/2,0,9/2)$ to $(5/2,0,5/2)$, $(5/2,0,5/2)$ to $%
(1/2,1,7/2)$, and $(1/2,1,7/2)$ to $(3/2,0,11/2)$ where the total spin
changes with $|\Delta S_{z}|\geq 1$.

Experimentally, it is more convenient to tune the number of electrons\cite%
{tokura} occupying the CQR. Therefore, we calculate the addition energy as a
function of the electron number in strong, intermediate and weak coupling
regimes as shown in Fig. 3. For $r_{0}=$ 4.0 a$_{0}$ and $\Delta =$ 1.6 meV (%
$d=35.8$ \AA\ ) in Fig. 3(a), the two rings are strongly coupled acting as a
single one. For a single ring, the closed shells are formed\cite{lin} at $N=$%
2, 6, 10, ... resulting in peaks in the addition energy. From the inset we
notice that $I_{z}$ increases monotonously with increasing $N$ because all
the electrons are in the bonding states. When we reduce the ring radius to $%
r_{0}=$ 2.0 a$_{0}$ while keeping the same tunneling energy $\Delta =$ 1.6
meV as shown in Fig. 3(b), we find the CQRs entering the molecular type
phase for $N\geq 7$. The strong addition energy peak at $N=6$ indicates that
the system transfers from a closed shell ring atomic type phase $(0,0,3)$
into a molecular type phase $(1/2,0,5/2)$ for $N=7$ where there is one
electron in the lowest antibonding level $\epsilon _{0,0}$. From $N=7$ to $8$%
, $I_{z}$ changes from $5/2$ to $4$ as shown in the inset of Fig.
3(b) with a total isospin change $|\Delta I_{z}|=3/2>1/2$ resulting
in a strong reduced current peak \cite{bart,tokura} because of
isospin blockade. In this intermediate coupling regime, the ground
state phase depends strongly on both the ring radius and the
inter-ring distance as shown in Fig. 2. This is clearly reflected in
the addition energy. The peak in the addition energy at $N=12$ in
Fig. 3(b) corresponds to the molecular type phase $(0,0,4)$ where
there are ten paired electrons in the bonding levels and two paired
electrons in the antibonding level $\epsilon _{0,0}$.

When the ring radius is kept as a constant, the CQRs become weakly coupled
for large inter-ring distance. Such a situation is presented in Fig. 3(c)
for $r_{0}=$ 2.0 a$_{0}$ and $d=$67.4 \AA ($\Delta =$ 0.15 meV) where the
CQRs are in the weak-coupling regime. Because the system evolutes to two
separated rings at $d\rightarrow \infty $ ($\Delta \rightarrow$ 0), one
should find closed shells at $N=$4, 12, 20, ... for single rings. We see
these features in the inset in Fig. 3(c). However, we also find a pronounced
addition energy peak at $N$=8 indicating coupling of two half-filled rings
with total spin $S_{z}$=2. The peak at $N$=12 corresponds to two closed
shell rings of 6 electrons in each ring.

\section{Conclusions}
In summary, new quantum ring molecular type phases as well as ring atomic
type phases are predicted for the CQRs which can be probed experimentally
through current measurements. Our results show transitions of these
artificial molecular phases from the CQDs ($r_{0}=0$) to the CQRs and also
from two strongly coupled quantum ring AMs to their dissociation in the weak
tunneling regime. In comparison to the system of CQDs, quantum tunneling
effects are enhanced in the present system where the AMs require less
tunneling energy to entering the molecular type phases.

\acknowledgments This work was supported by FAPESP and CNPq (Brazil)
and by the Flemish Science Foundation (FWO-Vl) and the Belgium
Science Policy. Part of this work was supported by the EU network of
excellence: SANDiE.

\newpage
\begin{center}
\Large\bf{Figures}
\end{center}
\begin{figure}[h!]
\begin{center}
\end{center}
\caption{(color on line) Electron energy levels $\protect\epsilon
_{n,m}$ in a single quantum ring as a function of the ring radius.
The potential height in the center of the ring is indicated by the
thick dot-dashed line.}
\end{figure}

\begin{figure}[h!]
\centering
\caption{(Color on line) The ground-state phase diagrams of the AMs
composed by CQRs in the plane of the inter-ring tunneling energy
(inter-ring distance) vs. the ring radius for (a) N=8, (b) N=9, (c)
N=11, and (d) N=13 electrons. $\hbar \protect\omega _{0}=5$ meV. The
grey (green) region indicates the phases which can be found at
$r_0\to $0 ($r_0\to \infty$). The yellow region shows the phases
which exist only for intermediate ring radius. The insets indicate
the single particle picture representation for each phase where the
bonding (antibonding) states are represented on the left (right) of
the vertical dotted line. }
\end{figure}

\begin{figure}[h]
\caption{The addition energy as a function of the number of electrons in the
CQRs with (a) $r_{0}=$ 4.0 a$_{0}$ and $\Delta =1.6$ meV; (b) $r_{0}=$ 2.0 a$%
_{0}$ and $\Delta =1.6$ meV; and (c) $r_{0}=$ 2.0 a$_{0}$ and $\Delta =0.15$
meV. The insets show the corresponding total spin $S_{z}$, angular momentum $%
M_{z}$ and isospin $I_{z}$. }
\end{figure}

\end{document}